\documentclass[aps,prl,twocolumn,showpacs,groupaddress]{revtex4}
\usepackage{amsmath}
\usepackage{amsfonts}
\usepackage{amssymb}
\usepackage{epsfig}
\usepackage{graphicx}
\newcommand{\be}{\begin{equation}}
\newcommand{\ee}{\end{equation}}
\renewcommand{\phi}{\varphi}

\begin{document}

\title{Nonequilibrium glassy dynamics of self-propelled hard disks}

\author{Ludovic Berthier}
\affiliation{Laboratoire Charles Coulomb, UMR 5221, CNRS and Universit\'e
Montpellier 2, Montpellier, France}

\date{\today}

\begin{abstract}
We analyse the collective dynamics of self-propelled 
particles in the large density regime where passive particles 
undergo a kinetic arrest to an amorphous glassy state.
We capture the competition between self-propulsion and crowding effects
using a two-dimensional model of self-propelled hard disks, 
which we study using Monte-Carlo simulations.
Although the activity drives the system far from equilibrium, 
self-propelled particles undergo a kinetic arrest, which 
we characterize in detail and compare with its equilibrium 
counterpart. In particular, the critical density for dynamic arrest 
continuously shifts to larger density with increasing activity, 
and the relaxation time is surprisingly well described by an 
algebraic divergence resulting from the emergence of highly collective 
dynamics. These results show that dense assemblies of active particles 
undergo a nonequilibrium glass transition which is profoundly 
affected by self-propulsion mechanisms.
\end{abstract}





\maketitle

The equilibrium physics of dense particle systems 
is usually understood in the framework of statistical mechanics because it 
stems from the competition between particle interactions
and thermal fluctuations~\cite{hansen}. In particular, phase transitions 
towards crystalline or amorphous structures are routinely observed 
at equilibrium~\cite{tabor}. This approach is challenged for particle 
assemblies that are not uniquely driven by thermal fluctuations, but 
can also pump energy from their environment to self-propell 
themselves~\cite{review1,review2}. Active particles are presently the focus 
of a large interest, fueled by experimental developments 
allowing the study of both natural living systems
(such as bacteria~\cite{bacteria} and cells~\cite{cells}) 
and synthetic colloidal~\cite{colloid} and
granular~\cite{granular} particles. It is thus important to understand
if and how equilibrium phenomena are affected by this novel 
type of nonequilibrium driving and dissipation mechanisms.

We study the behaviour of self-propelled particles when steric effects 
compete with self-propulsion~\cite{silke,sperl,jorge}.
Provided crystallization is suppressed (for instance by
size polydispersity), simple
fluids at thermal equilibrium display at large 
density a gradual transformation towards an arrested disordered 
state~\cite{rmp}. While not yet systematically explored,
this situation is of experimental interest for 
several systems of active particles. For instance,
the complex mechanical properties of epithelium tissues result 
from the influence of self-propulsion mechanisms for close-packed 
cells~\cite{cells,tissue}, while dense 
bacterial colonies are being studied 
experimentally~\cite{densecells}. Self-propelled colloidal
and granular assemblies can also be compressed 
to large densities~\cite{densecoll}.
On the theoretical side, it was recently suggested 
that active particles, despite being far from 
equilibrium, could display kinetic arrest with qualitative 
analogies, but also strong differences, with the 
equilibrium glass transition~\cite{jorge}. This suggestion,
obtained in the framework of mean-field approaches to driven glassy
dynamics, is by no means obvious as slow dynamics 
is usually fully disrupted by driving forces, such as 
a shear flow~\cite{sgr,BBK}. Therefore, it is important 
to study whether the competition between particle scale driving forces and 
glassy dynamics can yield a nonequilibrium phase transition even 
in a more realistic situation, which is our
central goal.

To this end, we seek a minimal model to study the impact of self-propulsion 
on the dynamics of dense assemblies of self-propelled particles, 
allowing us to interpolate smoothly between the well-known (but already 
complex) equilibrium glassy dynamics, and the driven active case. 
Therefore, by contrast with detailed numerical study of active matter
at moderate densities, our model incorporates active motion
following the simplest models of active matter,
neglecting for instance hydrodynamic interactions, particle anisotropy
or aligning interactions.
We work in two spatial dimensions, which is experimentally 
relevant~\cite{tissue,granular} and typically preferred in earlier 
studies~\cite{review2,silke,active_num}.  
To capture crowding effects, we use a 50:50 binary 
mixture of hard disks with diameter ratio 
$\sigma_1/\sigma_2 = 1.4$, which both suppresses 
crystallization and displays realistic glassy dynamics at equilibrium.
The hard sphere model is also convenient because it does not require the 
introduction of an energy (or a temperature) scale. Instead it is
uniquely controlled, at equilibrium, by the packing fraction, 
$\varphi = \pi N ( \sigma_1^2 + \sigma_2^2) / (2 L^2)$ 
for $N$ particles in a system of linear size $L$, using periodic
boundary conditions. We express lengthscales in units of $\sigma_1$. 

We use off-lattice Monte-Carlo simulations to study the glassy dynamics
of the model~\cite{demian}. At equilibrium, an elementary move proceeds 
as follows. At time $t$, a particle is chosen at random, say particle $i$, 
and a small random displacement $\vec{\delta_i}(t) = \delta_0 \vec{\xi_i}(t)$ 
is proposed, where $\delta_0$ sets the typical amplitude of the 
moves, and $\vec{\xi_i}(t)$ is a random vector drawn 
independently at each step from a unit square centered around the 
origin with a flat distribution. 
The move is accepted provided it creates no overlap 
with another hard disk. In equilibrium conditions, it was established
that Monte-Carlo simulations meaningfully and efficiently
describe the slow dynamics of glass-formers provided the 
jump length $\delta_0$ is adjusted by seeking a compromise between 
a small value (where the method becomes equivalent to 
Langevin dynamics) and a large value (creating unphysical 
non-local moves)~\cite{berthier-kob}. We use $\delta_0=0.1 
\sigma_1$, so that $\delta_0$ does not influence
the physics, apart from a trivial rescaling of the time.
We have explicitely checked that our results are not qualitatively 
affected by this choice. 
Timescales are expressed in Monte-Carlo steps, such that 
one time unit $\tau_{MC}$ represents $N$ attempted particle moves. 
The equilibrium dynamics of the hard disk system is 
thus characterized by a unique 
control parameter, the packing fraction $\phi$.

Following previous work~\cite{silke,selfprop},
we introduce a self-propulsion mechanism using a persistence timescale,  
$\tau$, defined as a finite timescale governing rotational
diffusion so that our model falls into the class of 
`apolar active particles', characterized in particular by the absence
of any alignement rule. Rotational diffusion is easily implemented
in the Monte-Carlo algorithm by generating time correlated 
random displacements. In practice, we initialize $\vec{\delta_i}(t=0)$ 
as before, $\vec{\delta_i}(0) = \delta_0 \vec{\xi_i}(0)$, 
but introduce temporal correlations between successive attempted displacements 
at times $t$ and $t'$:
\be 
\vec{\delta_i}(t) =  \vec{\delta_i}(t') + \delta_1 \vec{\xi_i}(t),
\label{markov}
\ee 
constraining $|\delta_{i,\alpha}(t)| \leq \delta_0$, 
and $\delta_1 \leq \delta_0$. 
As in equilibrium, the particle move is only accepted if it creates 
no overlap between particles, but the random displacement is updated 
as in Eq.~(\ref{markov}) independently of the acceptance condition,
thus generating a fixed persistent time $\tau$ for the 
orientation. Equation (\ref{markov}) means that particle displacements 
have the same amplitude 
as in equilibrium, but now keep a memory of previous displacements 
over a finite timescale, $\tau = (\delta_0 / \delta_1)^2$
(expressed in Monte-Carlo time units defined above).
Equation (\ref{markov}) represents a discrete-time analog 
of the Langevin dynamics studied in Refs.~\cite{silke,selfprop}, which is 
recovered in the limits $\delta_0, \delta_1 \to 0$, keeping 
the persistence time fixed~\cite{demian}. 
Self-propulsion is thus uniquely characterized 
by $\tau$, which reduces, in the dilute limit,
to the persistence time of a persistent random walk 
motion. Equivalently, this control parameter $\tau / \tau_{MC}$ 
can be seen as an adimensional 
rotational P\'eclet number~\cite{Gompperpeclet}. Because thermal fluctuations 
only affect rotational degrees of freedom, 
the translational P\'eclet number is not a convenient 
control parameter in our model~\cite{selfprop}.

While clearly minimal, the model efficiently 
captures the competition between steric hindrance 
(controlled by $\varphi$) and self-propulsion
(controlled by $\tau$). We performed extensive simulations in 
the steady state varying 
$(\phi, \tau)$ over a broad range, typically using $N=10^3$ particles.
Our longer simulations last $10^{10}$ steps. The model 
is presented more extensively and compared to alternative numerical
models in Ref.~\cite{demian}, which shows in particular that the system remains 
homogeneous at all densities, in contrast with earlier 
numerical works~\cite{hagan,marchetti,bialke,gompper,footnote}. 
Here we concentrate on the large density regime, not explored before. 

\begin{figure}
\psfig{file=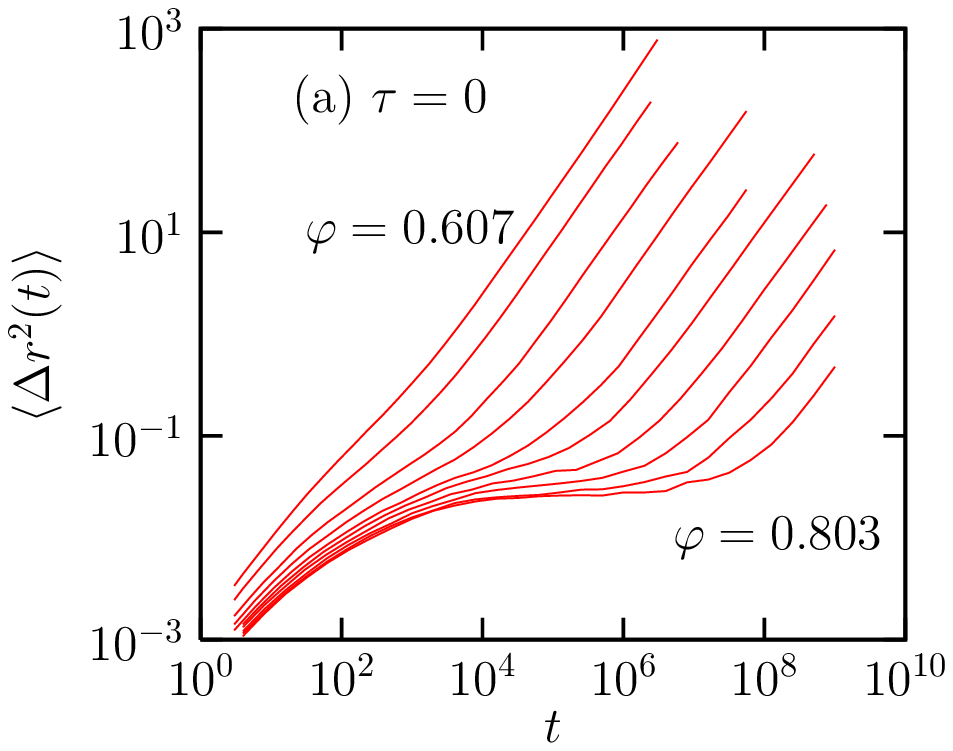,width=4.25cm}
\psfig{file=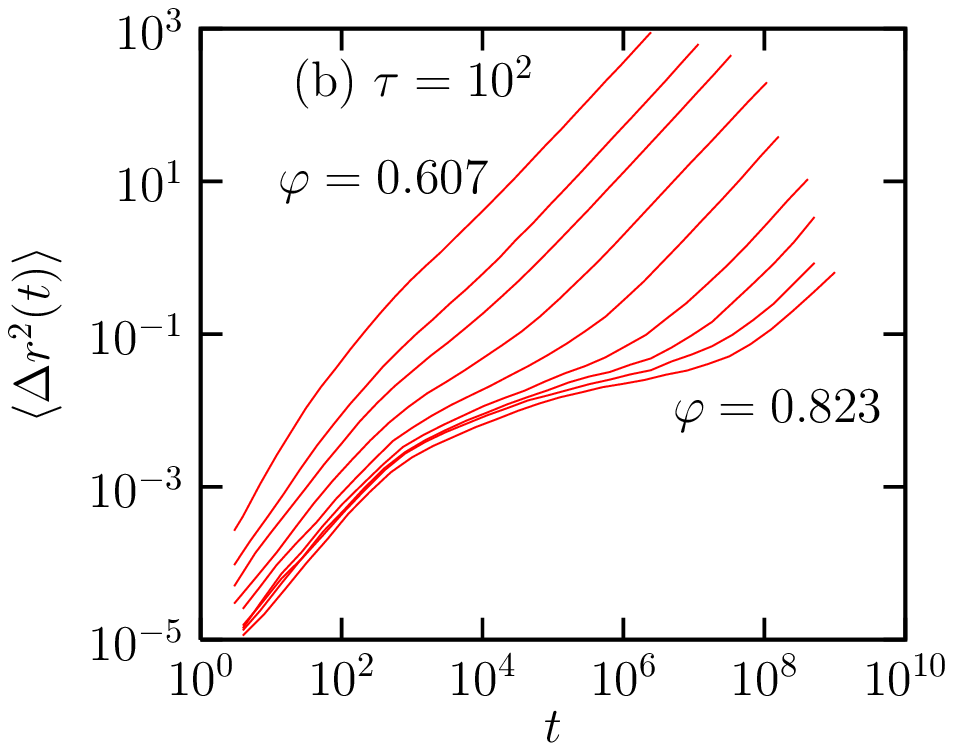,width=4.25cm}
\psfig{file=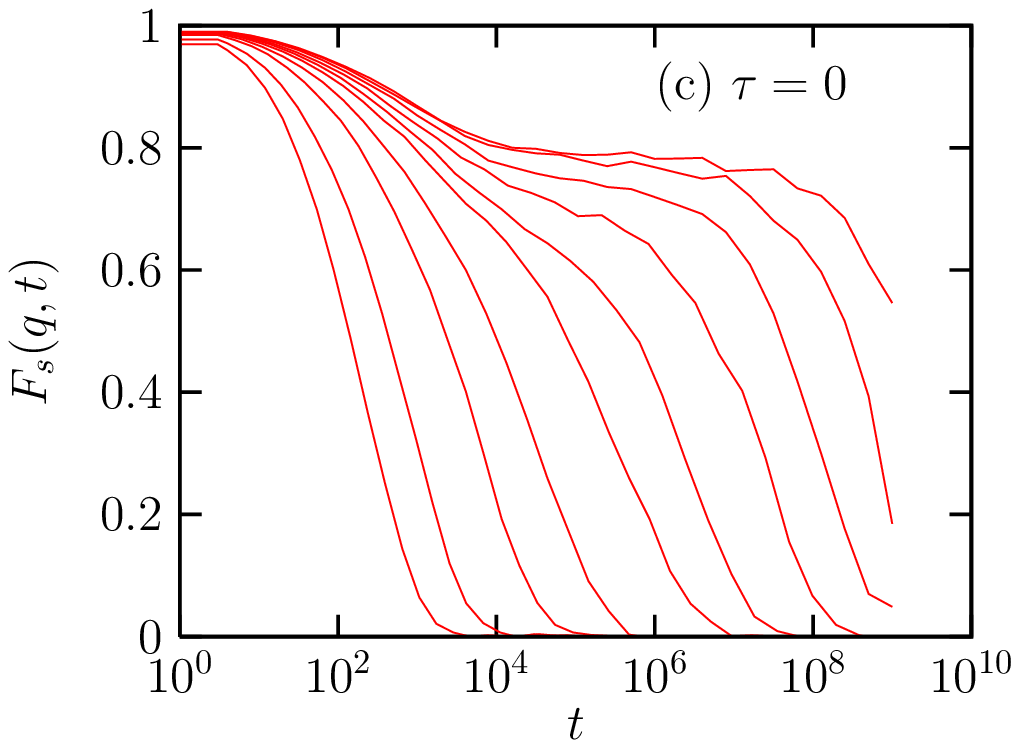,width=4.25cm}
\psfig{file=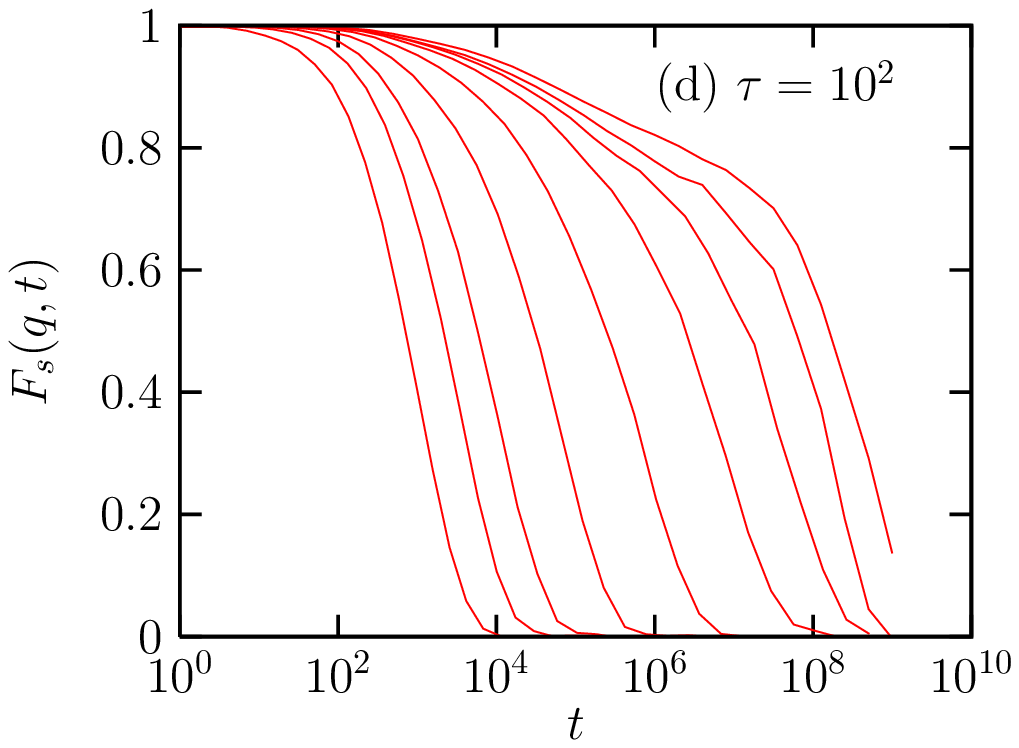,width=4.25cm}
\caption{\label{fig1}
Glassy dynamics for (a,c) equilibrium and
(b,d) self-propelled hard disks with 
persistence time $\tau=10^2$. 
Time dependence of the mean-squared displacement, 
Eq.~(\ref{msd}), and the self-intermediate scattering function, 
Eq.~(\ref{isf}) for increasing packing fraction.
From left to right in (a, c): $\phi=0.607$, 0.700, 0.754,
0.773, 0.785, 0.790, 0.795, 0.800, 0.802, and 0.803.   
From left to right in (b, d): $\phi=0.607$, 0.700, 0.743, 
0.781, 0.806, 0.819, 0.823, 0.825, and 0.828.  
Note the change of vertical scale between (a) and (b). 
Two-step, glassy dynamics emerge in both cases, suggesting that self-propelled 
particles undergo a nonequilibrium glass transition.}
\end{figure}

We start our analysis with a brief description of the 
glassy dynamics observed when $\phi$ increases in the absence of 
self-propulsion, $\tau=0$. In Fig.~\ref{fig1}(a) we show the time 
dependence of the mean-squared displacement, 
\be
\langle \Delta r^2(t) \rangle = 
\langle | \vec{r_j}(t) - \vec{r_j} (0) |^2 \rangle, 
\label{msd}
\ee
where $\vec{r_j}(t)$ denotes the position of particle $j$ at time $t$ and
brackets indicate an ensemble average performed in 
steady state conditions. The average is specialized 
to large particles, the slowest 
component of the binary mixture. While particles 
diffuse rapidly for moderate packing fractions, 
diffusion slows down dramatically as $\phi$ increases. We cannot 
observe long-time diffusion in the time window explored
by the simulation for $\phi > 0.803$, because 
it is too slow. Another signature of glassy dynamics is the emergence
of the intermediate time plateau in Fig.~\ref{fig1}(a), indicating 
that particle dynamics is essentially a `caged' motion
at intermediate times.
This two-step dynamics is confirmed in Fig.~\ref{fig1}(c) by the time evolution
of the self-intermediate scattering function,
\be 
F_s(q,t) = \langle e^{i \vec{q} \cdot [ \vec{r_j}(t) - 
\vec{r_j} (0) ]} \rangle,
\label{isf}
\ee 
which quantifies dynamics occurring over a 
length $2 \pi / |\vec{q}|$. 
We perform a circular average over wavevectors 
corresponding to the typical interparticle distance, 
$|\vec{q}| = 6.2$, corresponding to the first peak of the 
structure factor.

Turning to self-propelled particles
with $\tau=10^2$ in Figs.~\ref{fig1}(b,d), we find that dynamics 
again becomes slower as $\phi$ increases,
with the development of complex time dependences in both time correlators. 
Clear differences with the equilibrium situation already emerge for
moderate densities and short-times, where
active particles move ballistically as a direct result 
of self-propulsion.

At larger density, the
plateau in $F_s(q,t)$ is less pronounced for self-propelled than
equilibrium particles. Mean-squared displacements take
lower values at short-times,
showing that cage dynamics is profoundly affected by the 
particle activity. While a fast erratic exploration 
of the cage results from thermal noise, persistent motion 
is impossible within a cage. Instead, we observe that self-propelled 
particles transiently `stick' to the neighbor found in the 
direction of motion for a duration $\tau$, until randomization of the 
direction of motion allows further displacement.
As a result, particles can be fully arrested at short times,
reducing $\langle \Delta r^2(t) \rangle$ in this regime.
The cage exploration thus occurs over a broader distribution of 
times, which produces a complex time dependence of 
$F_s(q,t)$ and $\langle \Delta r^2(t) \rangle$ in the plateau regime.
Physically, thermal vibrations are suppressed by 
the persistent motion and occur over a time $\tau$ 
that may become decoupled from the microscopic scale. 
This observation is crucial, because the equilibrium physics of 
hard spheres is controlled by entropic forces~\cite{hansen}, which are then 
considerably impacted by self-propulsion. Finally,
although less mobile at short times, self-propelled
particles diffuse much faster at long times.
Diffusive motion is for instance still observed for $\phi = 0.823$
and $\tau=10^2$, while it is fully arrested 
at this density at equilibrium. These observations reveal 
that the nature of the glass transition is dramatically modified 
for active particles.

\begin{figure}
\psfig{file=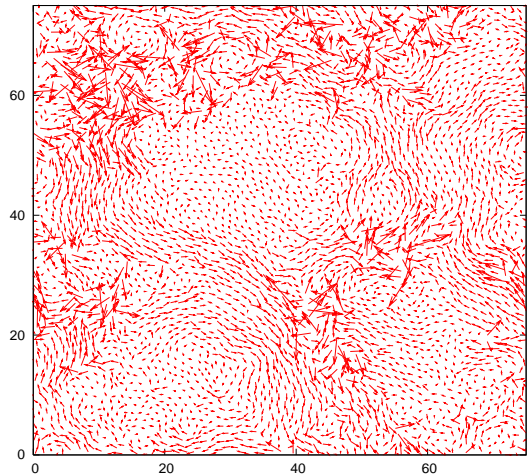,height=7.5cm,angle=-90,clip}
\caption{\label{dynhet} 
Displacement map for self-propelled disks
with $\phi=0.823$ and $\tau=10^2$ over a time 
$t \approx 1.5 \cdot 10^7$ corresponding to structural relaxation.
It shows the emergence of collective motion correlated  
over large distance in dense assemblies of active particles.}
\end{figure}

We show in Fig.~\ref{dynhet} a displacement map for 
self-propelled particles with $\tau=10^2$ and large density
$\phi = 0.823$, measured over a time interval corresponding
to structural relaxation (see below for a definition).
Clearly, the flow of self-propelled 
particles at large density is spatially correlated over
large distances, and thus displays large scale 
dynamic heterogeneity~\cite{book,physics}. Spatially correlated displacements
represent a form of emergent collective motion arising 
from the competition between self-propulsion and steric effects,
which differs qualitatively from earlier observations in active 
particle systems~\cite{collective}. The analogy between collective motion 
and dynamic heterogeneity in epithelium tissues was noted~\cite{tissue}. 

\begin{figure}
\psfig{file=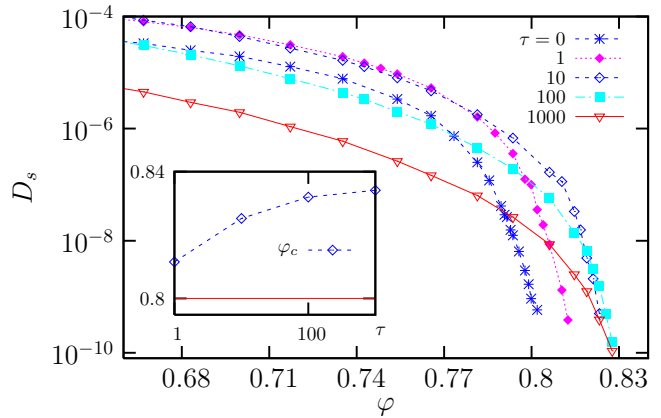,width=8.5cm}
\caption{\label{fig2}
Density dependence of the diffusion constant for 
different persistence time. 
Inset: The critical density $\phi_c$ obtained
from Eq.~(\ref{powerlaw}), increases continuously 
with $\tau$.}
\end{figure} 

We extract the long-time self-diffusion constant, $D_s$, from its definition, 
$D_s \equiv \lim_{t \to \infty}  \langle \Delta r^2(t) \rangle / (4t)$,  
and report in Fig.~\ref{fig2} the density evolution of 
$D_s$ for equilibrium and self-propelled disks.
These data confirm that in all cases $D_s$ 
decreases sharply upon increasing $\phi$, as it varies 
by nearly 6 orders of magnitude between the simple
fluid at $\phi \approx 0.6$ to the dense regime near 
$\phi \approx 0.8 - 0.83$. 
Increasing $\tau$ has two opposite effects, as demonstrated 
by the non-monotonic evolution of $D_s$ with $\tau$ at fixed 
$\phi$. First, increasing $\tau$ slows down diffusion as 
particles need to wait at least a timescale $\tau$ to see their orientation 
diffuse significantly. This effect dominates at moderate densities, 
where $D_s$ decreases with increasing $\tau$, see Fig.~\ref{fig2}.
However, self-propulsion has a 
less trivial effect at large $\phi$, where it accelerates the 
dynamics dramatically. For $\phi=0.8$, $D_s$ 
increases by 3 orders of magnitude between $\tau=0$ (equilibrium)
and $\tau=10$. Such an acceleration of the dynamics could result 
from the complete disappearance of the glass transition
(as for shear flow~\cite{BB}), but the data
in Fig.~\ref{fig2} suggest a different scenario. 
Although dramatically affected, the density 
dependence of the diffusion constant for $\tau >0$ 
remains very sharp, indicating that diffusion will cease
above a density which remains well-defined. In other words,
our simple model of self-propelled hard disks
displays a nonequilibrium form of dynamic arrest, despite the presence
of driving forces with finite amplitude. 
This finding is fully consistent with the theoretical 
suggestion in Ref.~\cite{jorge}.

We quantify the effect of the self-propulsion 
on the location of the glass transition by extracting 
a critical density $\phi_c$ using  a power law description: 
\be
D_s \sim ( \phi_c - \phi )^\gamma,
\label{powerlaw}
\ee
where the exponent $\gamma$ and the critical density
$\phi_c$ might depend on $\tau$. Equation (\ref{powerlaw}) 
is inspired by equilibrium studies of the glass transition~\cite{gotze}, 
and can be derived in the framework of mode-coupling approaches~\cite{jorge}.
The evolution with $\tau$ of the fitted $\phi_c$ shown in 
Fig.~\ref{fig2} shows that it increases 
continuously, departing from its equilibrium value as soon as 
a finite persistence time $\tau > 0$ is introduced.
This confirms that the `reentrant' evolution of the 
diffusion constant with $\tau$ results from the competition
between a growing $\phi_c$ (which accelerates 
dynamics at constant $\phi$) and suppressed short-time vibrations (which 
slows down dynamics).  
The shift of $\phi_c$ with $\tau$, 
although small in absolute value, in fact
represents a spectacular effect. With thermal fluctuations,
it is not possible to observe structural relaxation for 
$\varphi \approx 0.83$, which is instead observed 
when $\tau \geq 10$. This implies 
that by breaking detailed balance and going out of equilibrium, 
the system discovers dynamical pathways that are essentially closed 
at equilibrium.

A tentative analogy with equilibrium systems suggests 
a physical explanation to the observed shift 
of the glass transition density with activity. Because 
hard disks cannot cross, self-propulsion then generates
an `effective' attractive force between particles moving 
towards one another~\cite{julien}. 
Equilibrium studies of adhesive hard spheres showed 
that the glass transition density increases with the strength of the 
attraction~\cite{attractive,reentrant}, because the equilibrium 
structure at short lengthscale is modified. Although 
structural changes occur in our system, it remains to 
be understood whether a mapping from self-propelled hard spheres to
equilibrium adhesive particles is meaningful~\cite{julien}.


\begin{figure}
\psfig{file=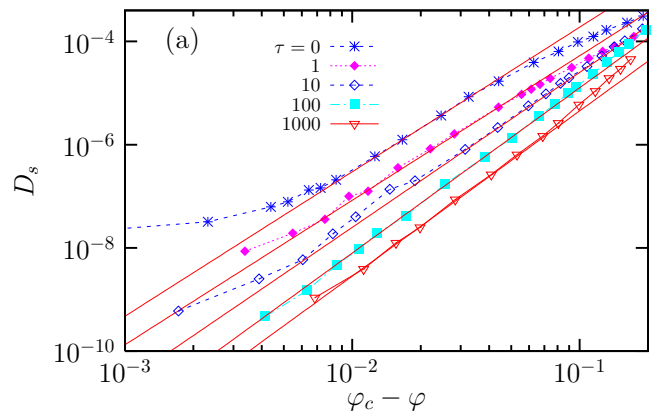,width=8.5cm}
\caption{\label{fig3} 
Critical representation of the diffusion constant showing that the 
range of validity of Eq.~(\ref{powerlaw}) increases 
with $\tau$ from about $2$ to $4$ decades of slowing down. } 
\end{figure}

The relaxation dynamics in self-propelled hard disks seems
however fundamentally
distinct from the equilibrium case. In the hard sphere fluid,
the onset of dynamic slowdown is described by 
a mode-coupling regime where Eq.(\ref{powerlaw}) holds, 
followed by a crossover to another regime controlled by activated 
relaxation events between low-lying metastable states~\cite{rmp,gio}. 
Therefore, introducing self-propulsion 
could affect the relevance of such activated dynamical processes. 
In Fig.~\ref{fig3} we confirm that the domain 
of validity of the power law in Eq.~(\ref{powerlaw}) 
increases from 2 to 4 decades between equilibrium and self-propelled 
particles with $\tau>10$. This suggests that mean-field,
mode-coupling types of approaches might represent a valuable 
theoretical starting point to describe the microscopic dynamics
of dense assemblies of active particles~\cite{jorge,sperl}. 

In conclusion, we found that self-propelled particles
undergo a nonequilibrium form of a glass transition at large density
that is distinct from its equilibrium counterpart, and characterized
by the emergence of a new form of collective motion directly
resulting from the interplay between activity and steric effects.

\acknowledgments

While completing this manuscript, R. Ni kindly sent a preprint 
reporting Brownian dynamics simulations of a different model of 
self-propelled hard spheres where the glass transition shifts
with activity~\cite{ni}. 
I also thank A. Ikeda, D. Levis and G. Szamel for discussions.
The research leading to these results has received funding
from the European Research Council under the European Union's Seventh
Framework Programme (FP7/2007-2013) / ERC Grant agreement No 306845.


\begin{thebibliography}{99}
\bibitem{reentrant}
For attractive forces, dynamics slows down in the limit of 
large attractive forces because gelation intervenes when long-lived bonds
form between particles. This causes a reentrant 
glass transition~\cite{attractive}. In our case, the persistence time fixes
an upper limit for the `effective' bond lifetime formed between 
self-propelled particles moving toward one another.
\end{thebibliography}
\end{document}